\documentclass{article}

\usepackage{arxiv}

\usepackage[utf8]{inputenc} 
\usepackage[T1]{fontenc}    
\usepackage{hyperref}       
\usepackage{url}            
\usepackage{booktabs}       
\usepackage{amsfonts}       
\usepackage{nicefrac}       
\usepackage{microtype}      
\usepackage{lipsum}
\usepackage{graphicx}
\graphicspath{ {./} }

\title{Improving VANET Simulation Channel Model in an Urban Environment via Calibration Using Real-World Communication Data}

\author{
 Ahmed Gammaa \\
 Center for Urban Informatics and Progress\\
  University of Tennessee at Chattanooga\\
  Chattanooga, TN, USA, 37403\\
  \texttt{wvt953@mocs.utc.edu} \\
   \And
 Seyedmehdi Khaleghian \\
 Center for Urban Informatics and Progress\\
  University of Tennessee at Chattanooga\\
  Chattanooga, TN, USA, 37403\\
  \texttt{kpk628@mocs.utc.edu} \\
  \And
 Toan Tran \\
  Emory University\\
  Atlanta, GA, USA \\
  \texttt{toan.tranviet@ieee.org} \\
  \And
  Mina Sartipi \\
 Center for Urban Informatics and Progress\\
  University of Tennessee at Chattanooga\\
  Chattanooga, TN, USA, 37403\\
  \texttt{Mina-Sartipi@utc.edu} \\
}

\begin{document}
\maketitle
\begin{abstract}
Wireless communication channels in Vehicular Ad-hoc NETworks (VANETs)  suffer from  packet losses, which severely influences the performance of their applications. There are several reasons for this loss, including but not limited to signal interference with itself after being reflected from the ground and other objects, the doppler effect caused by the speed of the vehicle, and buildings and other vehicles blocking the signal. As a result, VANET simulators must be calibrated in order to mimic the behavior of real-world vehicular communication channels effectively. In this paper, we calibrated an OMNET++(Objective Modular Network Testbed in C++)/Veins simulator for VANET's dedicated short-range communications (DSRC) protocol using the field data from the urban testbed in Downtown Chattanooga, TN. Channel propagation models, as well as physical layer parameters, were calibrated using a Genetic Algorithm (GA). The performance of the calibrated simulator was improved significantly in comparison with the default settings in Veins. The final results were compared to the real-world data collected from the testbed and performance shows that the final calibrated channel model performs better than uncalibrated models in simulating the packet delivery pattern of DSRC channels.
 
\end{abstract}


\section{I Introduction}

Dedicated Short Range Communications (DSRC) spectrum with a width of 75 MHz, or around 5.9 GHz, was assigned by the Federal Communications Commission (FCC) of the United States (and similar agencies in other countries) for use in VANETs \cite{federal2002amendment}. The IEEE has developed a pair of protocol standards (IEEE 802.11 and IEEE 1609) known as Wireless Access in Vehicular Environments (WAVE) to support DSRC \cite{ieee2010802, 6998915}. Since then, VANETs research has significantly improved transportation systems' efficiency and safety. VANET's vehicle-to-vehicle (V2V) and vehicle-to-infrastructure (V2I) communications offer driver assistance, road safety, traffic management, and infotainment applications.

VANETs use wireless communication at the network's edge, making them connectivity-sensitive and influence their performance in both urban areas and highways. Due to the density, congestion, and speed of vehicles in urban areas, a considerable chance of disconnection exists. Highways are unpredictable due to varying traffic and a small population. This way, VANET applications (especially safety-critical ones) must be validated and verified before deployment. Testing is done utilizing real-world trials via testbeds and through VANET simulators.
However, simulations are poor approximations of physical reality and testbed trials are unreproducible and prone to errors. Implementation of VANET applications is an iterative process of simulation, real world measurements, and calibration of simulator's parameters to get channel performance measures that closely follow reality. By calibrating their models, VANET simulators can simulate a realistic environment while operating at a high enough level of performance to study networks made up of thousands of connected vehicles.

In this paper, we incrementally built a channel model for a Veins simulator starting from a simple ideal free space model. We then moved towards a complex model that uses Lognormal slow fading model and Nakagami fast fading model. This complex pathloss model is not currently supported in INET (the library Veins relies on for wireless communication models). The new channel model parameters as well as physical layer parameter (transmission power and bitrate) were calibrated utilizing real-world data collected from the  MLK Smart Corridor, the urban testbed in Chattanooga, TN. Calibration was done through a Genetic Algorithm (GA) that produced the optimal set of parameters after running the simulation more than 5,500 times.

The remainder of this paper is structured as follows: background and related work are done in Section II. Section III summarizes the methodology we followed in this study. Section IV summarizes the experiment. Section V presents the results and key findings. Section VI concludes the paper with recommendations for future research.

\section{II Background and Related Work}
\subsection{VANET Architecture}

VANETs in their simplest form consist of two kinds of nodes, vehicles and Roadside Units (RSUs). Each vehicle is equipped with multiple devices, an Onboard Unit (OBU), one or more Application Units (AUs), and multiple sensors like Global Positioning System (GPS) for information collection. 

\subsubsection{On Board Unit}
The OBU is a WAVE device used by the vehicle to exchange messages with other OBUs and RSUs through a wireless communication. It is made up of a CPU, memory, and network interface for short-range wireless communication using IEEE 802.11p radio technology. The OBU provides multiple services for the vehicle. It connects the vehicle with OBUs and RSUs. It forwards messages from other OBUs (multip-hop communication). It also provides connectivity to the AUs.

\subsubsection{Application Unit}
The AU is a device that consumes VANET applications (protocols implemented and hosted by OBUs and/or RSUs called providers while consumed by other RSUs/OBUs/AUs called consumers). The AU connects to the provider or the internet using the communications capabilities of the OBU. A typical example is personal digital assistant (PDA) \cite{al2014comprehensive}.

\subsubsection{Roadside Unit}
RSUs represents routers connecting vehicles and the AUs and the internet. RSUs can also work as repositories that buffer information received from OBUs, other RSUs , and roadside sensors and cameras. A typical RSU has two network interfaces, one for connecting to the central infrastructure and the other for IEEE 802.11p wireless communications.

\section{The MLK Smart Corridor Testbed}
\begin{figure}[!ht]
  \centering
  \includegraphics[width=0.75\textwidth]{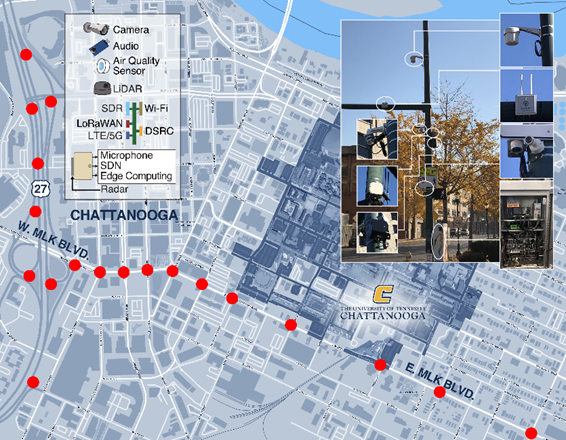}
  \caption{Map of the testbed in Downtown Chattanooga, TN. Capabilities of the testbed including the sensors, wireless communication capabilities, and computing resources are shown in the box on the left corner. Actual pictures of the testbed are shown on the right. Location of the dual-unit (DSRC and C-V2X RSUs) are marked on the map}\label{fig:mlk_corridor}
\end{figure}

MLK Smart Corridor is a smart city and connected vehicle testbed situated in Chattanooga, Tennessee's downtown as shown in Figure \ref{fig:mlk_corridor}. Eleven signalized junctions and 8 interstate poles are equipped with a range of communications, computer, and sensor equipment \cite{9006382}. A fiber network connects the infrastructure to an on-site data center at UTC. The testbed is backed by an event-driven data integration service that makes data collected on the MLK Smart Corridor available in real-time. Vehicle and pedestrian movement (speed, location, lane, size), signal phasing and timing (SPaT), connected vehicle data (BSM, TIM, PSM), and signal performance measurements (SPM) are examples of data sources. In addition, environmental sensors provide measurements of weather and particle matter at 1 Hz. A library and an application programming interface (API) provide access to historical and real-time data, respectively.

\section{Architecture of VANET simulator}

The primary components of a typical VANET simulator are mobility, network, and VANET simulators.

\subsection{Mobility Simulator}

Mobility simulators are primarily used to generate vehicle movement patterns under specified macro- and micro-mobility constraints. The macroscopic characteristics of vehicular traffic, such as road topology, speed restrictions, and safety requirements, are referred to as macro-mobility. Micro-mobility captures individual driver behavior such as acceleration, deceleration, as well as overall driving attitude. SUMO \cite{lim2017sumo} is one of the popular open-source mobility simulators that we have used in this paper.

\subsubsection{SUMO}
Simulating Urban Mobility (SUMO) is an open source, highly portable, microscopic and continuous traffic simulation tool designed to manage large networks of traffic. It has a wide range of capabilities for creating scenarios, including a comprehensive collection of tools for intermodal simulation~\cite{SUMO2018}. Multimodal traffic can be simulated using SUMO, which can import large-scale real-world maps from mapping platforms like OpenStreetMap (OSM), and can be interacted with using a socket-based communication interface called TraCI~\cite{krajzewicz2012recent}. Each vehicle's dynamics and behavior are separately represented at the microscopic level. Every vehicle on the roadway is simulated using "microscopic" models, which assume that the vehicle's physical capacities to move and the driver's controlling behavior influence the vehicle's behavior~\cite{lopez2018microscopic}. 

\subsection{Network Simulator}
To model the transmission of messages between linked nodes, a network simulator is utilized. This typically involves wired and, more importantly for VANETs, wireless communications. The nodes in VANETs are typically vehicles and RSUs. The networks simulator serves as a meduim where the VANET simulation runs. OMNeT++\cite{varga2010omnet++}, NS3\cite{carneiro2010ns}, and NS2\cite{issariyakul2009introduction} are examples of network simulators that are commonly used in VANETs.

\subsubsection{OMNET++}

OMNeT++ \cite{varga2010omnet++, homepage2022omnet++} is a C++-based discrete event simulator for modeling communication networks, multiprocessors and other distributed or parallel systems. OMNeT++ is public-source, and can be used under the Academic Public License that makes the software free for non-profit use. Instead of directly offering simulation components, OMNeT++ takes a framework approach by giving the essential machinery and tools to develop simulations. Various separately designed simulation models and frameworks, like the INET Framework \cite{meszaros2019inet}, cover certain application areas.

\subsection{VANET Simulator}
VANET simulator a framework that combines the mobility and networks simulators. The main objectives of the VANET simulator includes: (1) Defining the nodes of VANET (vehicles, RSUs, antennas, and other devices) as modules of the network simulator. (2) Interacting with the mobility simulator at run time to update the positions, speeds, and headings of the moving nodes of the simulation. (3) Implementing the WAVE standard and its protocols. There are many VANET simulators available as open-source (such as Veins, Eclipse MOSAIC, and VENTOS) as well as commercial ones like NetSim and EstiNet.

\subsubsection{Veins}
\begin{figure}[!ht]
  \centering
  \includegraphics[width=0.6\textwidth]{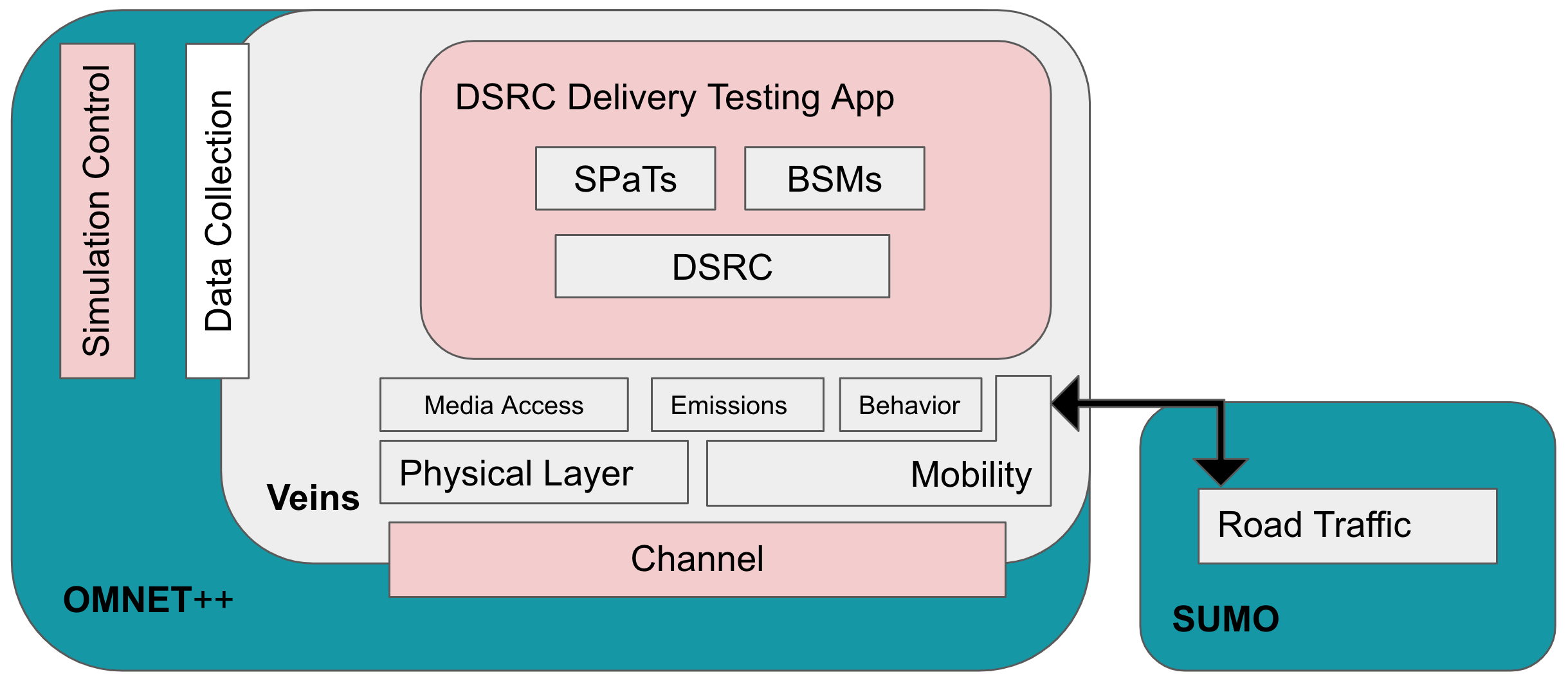}
  \caption{Architecture of Veins simulation platform. Adopted from \cite{weber2021vanet}}\label{fig:arch}
\end{figure}

Veins \cite{sommer2019veins} is a free and open-source framework for simulating vehicular networks. It is built with OMNeT++ and SUMO. Figure \ref{fig:arch} depicts the various modules that comprise the Veins architecture. Veins generates an OMNeT++ node for each vehicle in the simulation and then matches the node's motions to those of the vehicles in the road traffic simulator (i.e., SUMO). In this instance, both network and mobility simulations must run concurrently. This is enabled because a defined communication protocol, the Traffic Control Interface, enables bidirectional coupling (TraCI) \cite{wegener2008traci}. TraCI allows OMNeT++ and SUMO to exchange messages (such as mobility traces) via a TCP connection while the simulation is running \cite{sommer2010bidirectionally}.

\subsection{Simulation Channel Models Parameters Calibration}
Numerous studies were conducted to calibrate VANET simulators with real-world data. Adelin Miloslavov et al. \cite{miloslavov2011validation} implemented the Nakagami slow fading model for the NCTUns simulator \cite{wang2007design, wang2009nctuns} and used the NCTUns's built-in fast fading models.
The authors calibrated NCTUns using IntelliDrive SM Data Capture and Management Portal testbed data 
(open source data available online). Peng Su et al. \cite{su2016calibrating}
calibrated NCTUns simulator by collecting DSRC data using the Cooperative Vehicle Highway Testbed of the Saxton Transportation Operations Laboratory located in TFHRC in McLean using Latent Hypercube Sampling technique . Veins simulator was calibrated by Ioannis Mavromatis et al. \cite{mavromatis2017agile} using different scenarios (indoor, urban, suburban, and rural) with stationary On-board units (OBUs) throughout the operation.

\section{III Methodology}

\subsection{Channel Models for VANET Simulators}

All VANET simulators, including Veins, use models for large-scale path loss (slow fading) and small-scale (fast) fading to determine the received signal power and the decide whether a packet will be successfully received \cite{fall2012ns, rappaport1996wireless}. Veins depends on the well-known INET framework (INET is OMNeT++'s biggest collection of simulation models) \cite{meszaros2019inet}, which always employs the Free-space Model (FSM) for slow fading and several types of models for fast fading (ie; Lognormal, Nakagami, Reighleigh).

Since numerous studies \cite{yin2006dsrc, torrent2004broadcast} have found the Nakagami model to be the most suitable for VANETs, we adopted it in this study to simulate fast fading. The received signal strength after transmitting a signal with power \( x \) is given by the following equation \cite{4291825}:

\begin{equation}
    f_{R_{\ell}}(x; \Omega_{\ell}, m_{\ell}) = \frac{2m_{\ell}^{m_{\ell}}}{\Omega_{\ell}^{m_{\ell}}\Gamma(m_{\ell})} x^{2m_{\ell}-1} \exp\left(-\frac{m_{\ell}}{\Omega_{\ell}}x^{2}\right),
\end{equation}

where \( \Omega_{\ell} \) is the average power, and \( m_{\ell} \) is the Nakagami shape factor. The Nakagami model is superimposed on the Lognormal path-loss model, which is defined as follows \cite{fall2012ns}:

\begin{equation}
    \left[P_{r}(d)\right]_{dB} = \left[P_{r}(d_{0})\right]_{dB} + 10\alpha \log\left(\frac{d}{d_{0}}\right) + X_{dB}(\sigma_{dB}),
\end{equation}

where \( P_{r}(d) \) is the received signal power at distance \( d \), \( P_{r}(d_{0}) \) is the signal power at reference distance \( d_{0} \), \( \alpha \) is the free-space path-loss model (FSM) exponent, \( X_{dB}(\sigma_{dB}) \) is a normal distribution with zero mean, and \( \sigma_{dB} \) represents the standard deviation of the Lognormal fast fading model.

The reference signal power \( P_{r}(d_{0}) \) is determined using the FSM with the following equation \cite{fall2012ns}:

\begin{equation}
    P_{r}(d_{0}) = \frac{P_{t}G_{t}G_{r}\lambda^2}{(4\pi)^2 d_{0}^2 L},
\end{equation}

where \( P_{t} \) and \( P_{r} \) denote the transmitted and received power, respectively, \( \lambda \) is the wavelength, \( d_{0} \) is the reference distance, and \( L \) represents the system loss.

The final path-loss model used in this study is obtained by combining Equations (1) and (2).

\subsection{Genetic Algorithms for Parameters Calibration}

\begin{figure}[!ht]
  \centering
  \includegraphics[width=1\textwidth]{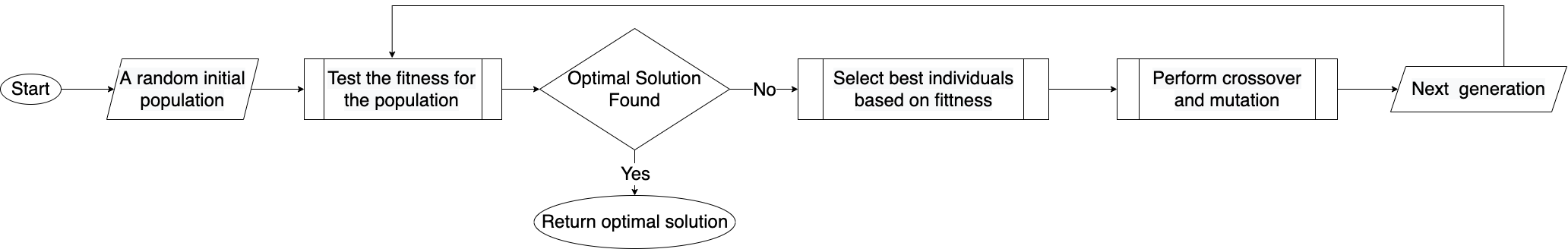}
  \caption{Flowchart of a Genetic Algorithm}\label{fig:ga_flow}
\end{figure}

Genetic algorithms (GA) are meta-heuristic optimization algorithms simulating the evolution of organisms in their natural environment \cite{mitchell1995genetic}. Given an optimization issue, a genetic algorithm uses evolutionary theory of selection (of the fittest) and mutation in order to simulate the development of solutions through time in order to locate an optimum or near-optimal solution. Figure \ref{fig:ga_flow} shows the steps of a GA optimization process.

\section{IV Experiment}

\subsection{Testbed Data Collection and Scenario}
\begin{figure*}[tp!]
        \centering
        \begin{minipage}[t]{.9\linewidth}
            \centering
            \includegraphics[width=\textwidth]{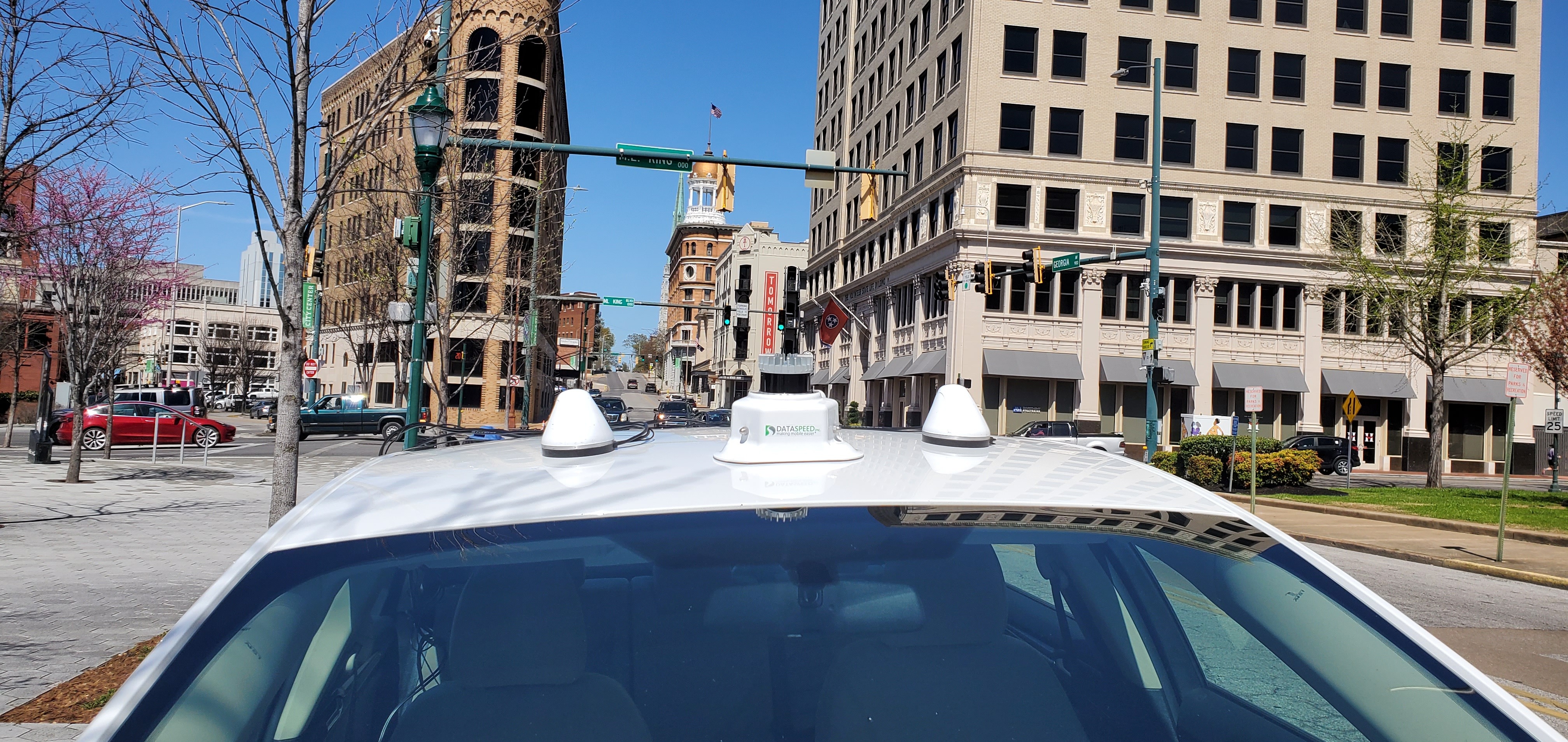}
        \end{minipage}
        \caption{The OBU-equipped vehicle used for data collection. }
        \label{fig:cuip_vehicle}
\end{figure*}

Center for Urban Informatics and Progress (CUIP) at the University of Tennessee at Chattanooga (UTC)  has outfitted a vehicle (2018 Ford Fusion) with an OBU that is wired to a 12 Volt power supply located in the vehicle's console. The vehicle also has a Global Positioning System (GPS) sensor, which is used to get real-time measurements on the location, speed, and heading of the vehicle at the precise moment of sending or receiving a message. The OBU exchanges messages with the RSUs through two antennae placed on the top center of the vehicle, roughly 24 inches apart. Figure \ref{fig:cuip_vehicle} shows the vehicle with its on board unit.

\begin{figure}[!ht]
  \centering
  \includegraphics[width=\textwidth]{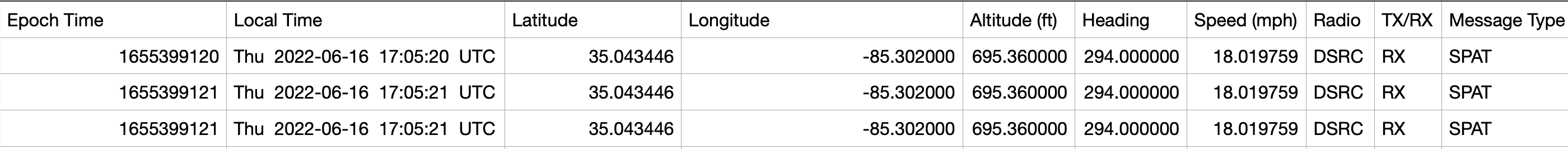}
  \caption{The collected data format}\label{fig:data_format}
\end{figure}

Throughout the scenario, the OBU equipped vehicle sent BSM messages to and Received SPaT messages from the MLK Smart Corridor RSU located at Georgina Ave and MLK Blvd intersection (the seventh RSU unit from the right on Figure~\ref{fig:mlk_corridor}). Data such as time, latitude, longitude, altitude (ft), heading, speed (mph), message transmission type (DSRC or C-V2X), message type (BSM or SPaT), and more are stored in the OBU memory as csv files at the time of sending and receiving of the two types of messages. We grabbed the csv files through the API and stored them in the library. Figure \ref{fig:data_format} depicts the format of collected data/

\subsection{Simulation Setup}

\begin{figure}[!ht]
  \centering
  \includegraphics[width=0.6\textwidth]{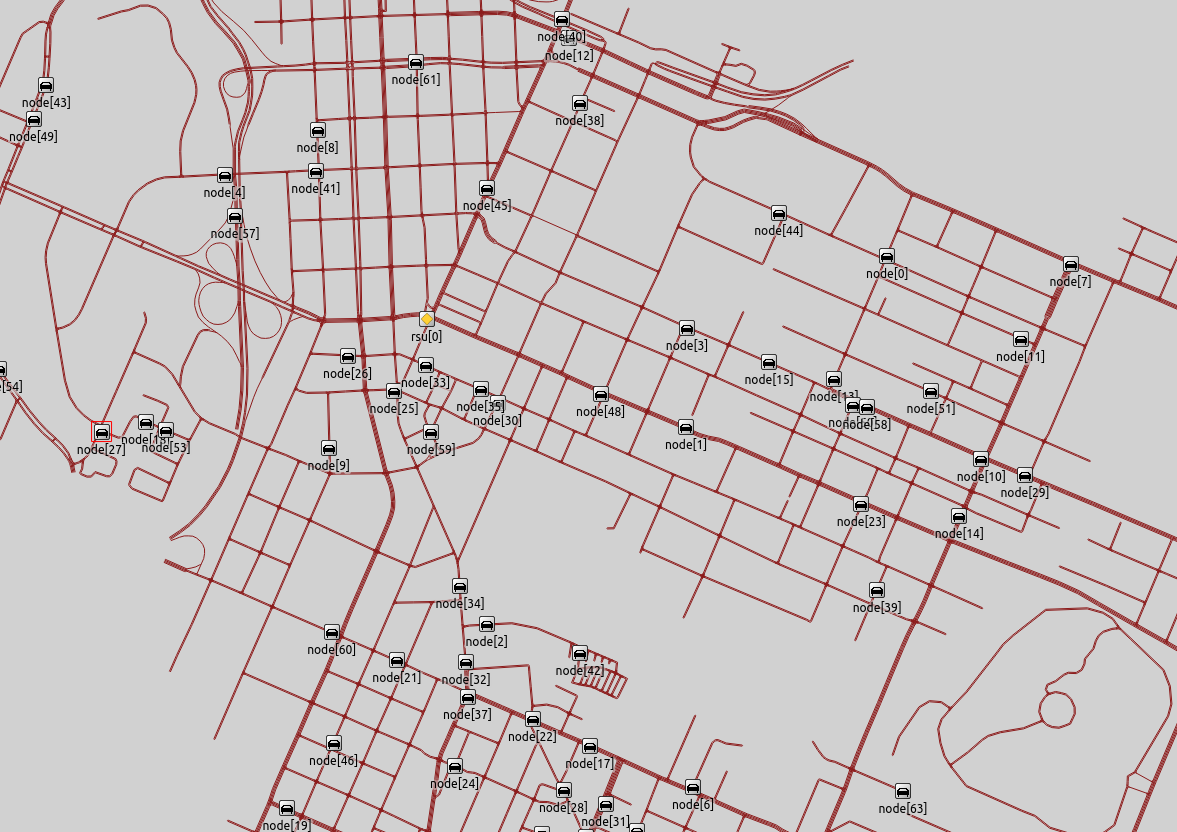}
  \caption{The Simulation Networks in the OMNET++ environment. The RSU is shown with yellow color.}\label{fig:sim_network}
\end{figure}

Under the OMNET++ environment, a Veins simulator was set up. A SUMO network simulating the same scenario is established (Figure \ref{fig:sim_network}). The SUMO's mobility is modified so that the target can be reached. The vehicle's path, average speed, stops, and accelerations are as close to the actual scenario as possible. The SUMO network and mobility files are imported into the OMNET++ environment, and the RSU is located by means of the OMNET++ configuration file (omnetpp.ini) along with a number of configurations. Veins is utilized to define the RSU and vehicle models as OMNET++ nodes. The TraCI server (SUMO) and network were both operational. 

For this study, there were a lot of changes made to the OMNET++/Veins/INET stack. The built-in definition of the BSM message was changed, and for the SPaT message, a new message type was made. Both messages were made to give the same kind of information as the testbed in Figure \ref{fig:data_format}. The RSU and vehicle definitions in Veins are changed so that they can send either BSM or SPaT and receive the other type. BSMs are sent from the OBU to the RSU, and SPaTs are sent from the RSU to the OBU. The INET path-loss definition was changed so that both large-scale (Nakagami) and small-scale (Lognormal) fading could be done on the same signal. Also, the definition was changed to return the decibel (dB) measure of signal attenuation, which is to be compared with the receiver's receiving sensitivity (in decibels).

\subsection{The Calibration Process}

First, the free-space pathloss model parameters (the exponent alpha and the system loss) were adjusted so that only negative pathloss values are generated (the maximum is zero at zero meters away). A severe penalty was associated with failing to meet the communication channel loss criterion of not having a positive path-loss value (the objective function returns 1000 RMSE for having a positive decibel loss). The purpose of this procedure is to prevent conceptual issues. For instance, the model can have a neat RMSE and a drop pattern that matches the testbed's while still causing some of the signals' power to increase (rather than decrease). This can hinder the model's ability to generalize to other scenarios.

The Lognormal model is then superimposed with the Nakagami model to create the slow fading and fast fading models, respectively. The average packet delivery ratio \cite{adrian2018study} per distance ranges was used to facilitate the calibration procedure. The root mean square error (RMSE) of the measurements between the collected and simulated data served as the genetic algorithm's objective function. The process of calibration is performed automatically by the genetic algorithm.

\begin{figure}[!ht]
  \centering
  \includegraphics[width=1\textwidth]{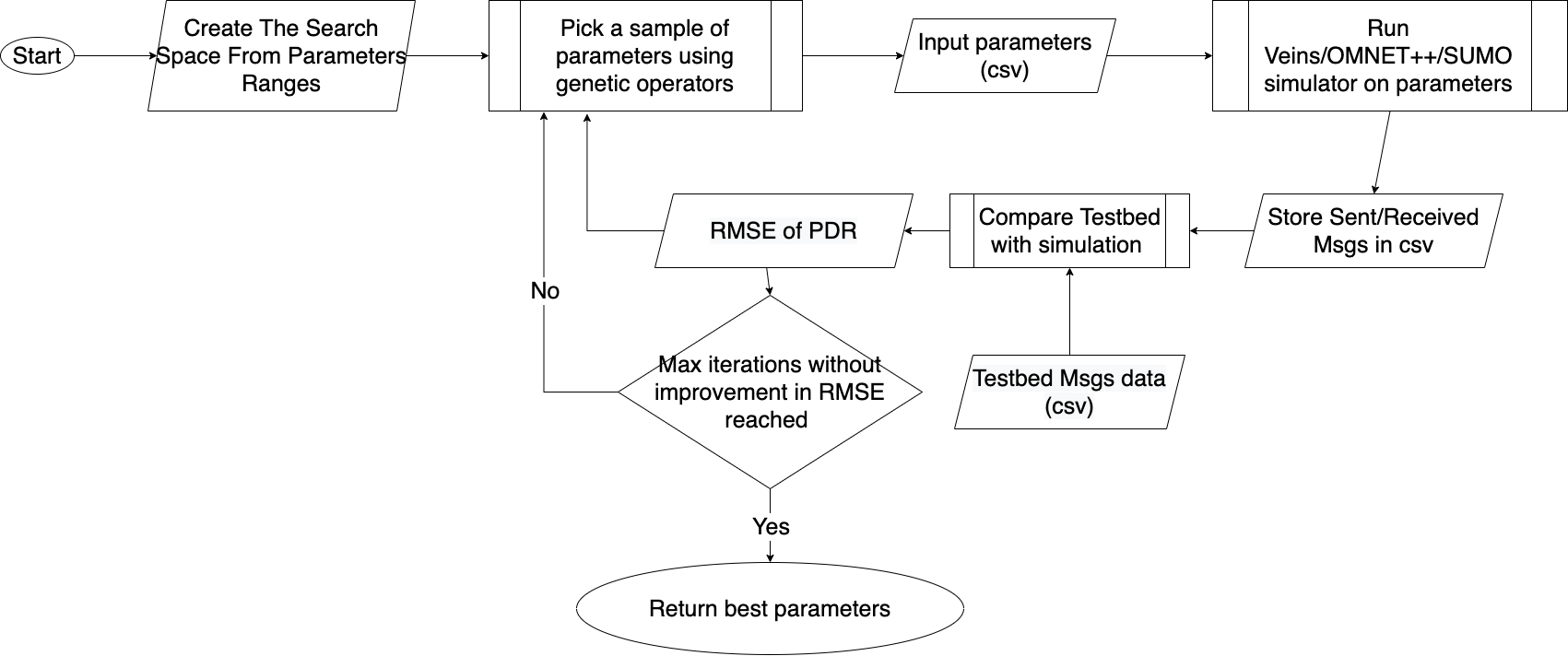}
  \caption{The calibration process using the GA optimization}\label{fig:calibration_flowchart}
\end{figure}

The Genetic Algorithm (GA) executed the simulation thousands of times in order to minimize the RMSE of packet delivery ratio. After each generation and based on the RMSE, our GA determines which parents to retain and which to replace for the subsequent generation as show in Figure \ref{fig:calibration_flowchart}. Following each step, the parameter set and associated RMSE were saved to a CSV file for analysis. The GA attempts to both converge on the global minimum and explore new routes to avoid local minimums. The simulation was run more than 5,500 times to determine the global optimal set of parameters.

\section{V Results and Findings}
\begin{figure*}[htp!]
\centering
             \begin{tabular}{@{}c@{}}
  \includegraphics[width=.55\textwidth]{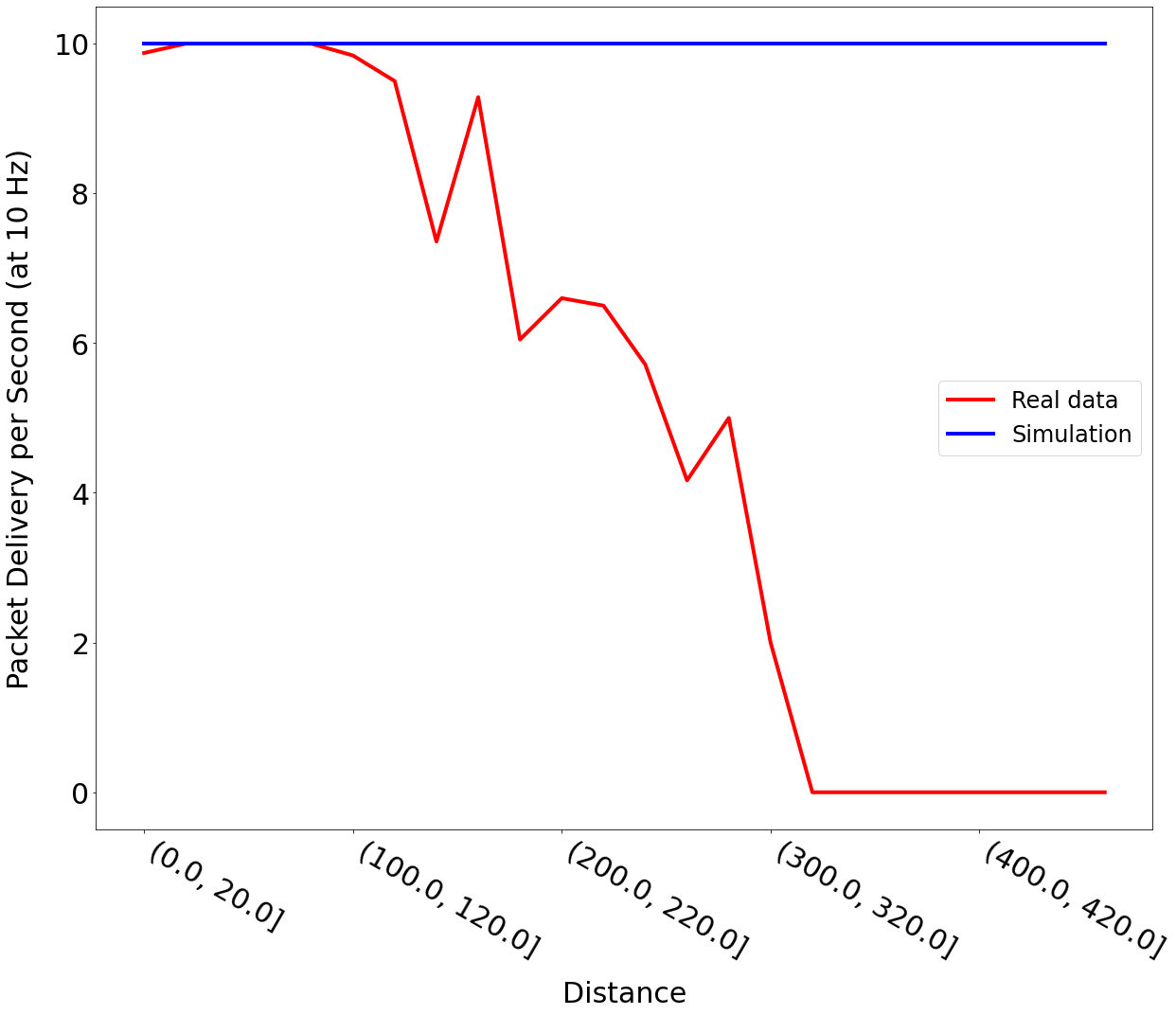}
             \end{tabular}
             \begin{tabular}{@{}c@{}}
  \includegraphics[width=.55\textwidth]{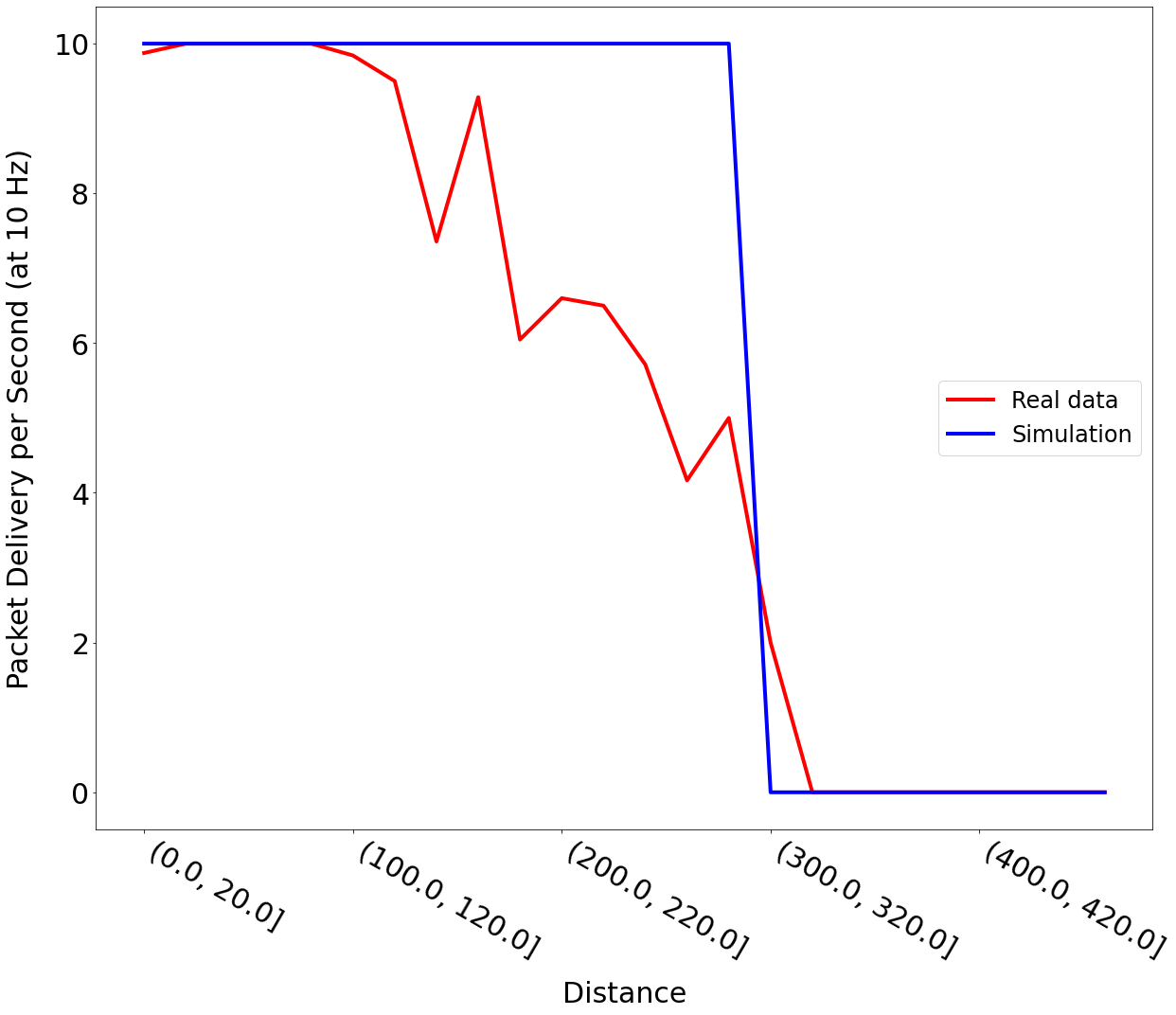}
             \end{tabular}
             \begin{tabular}{@{}c@{}}
  \includegraphics[width=.55\textwidth]{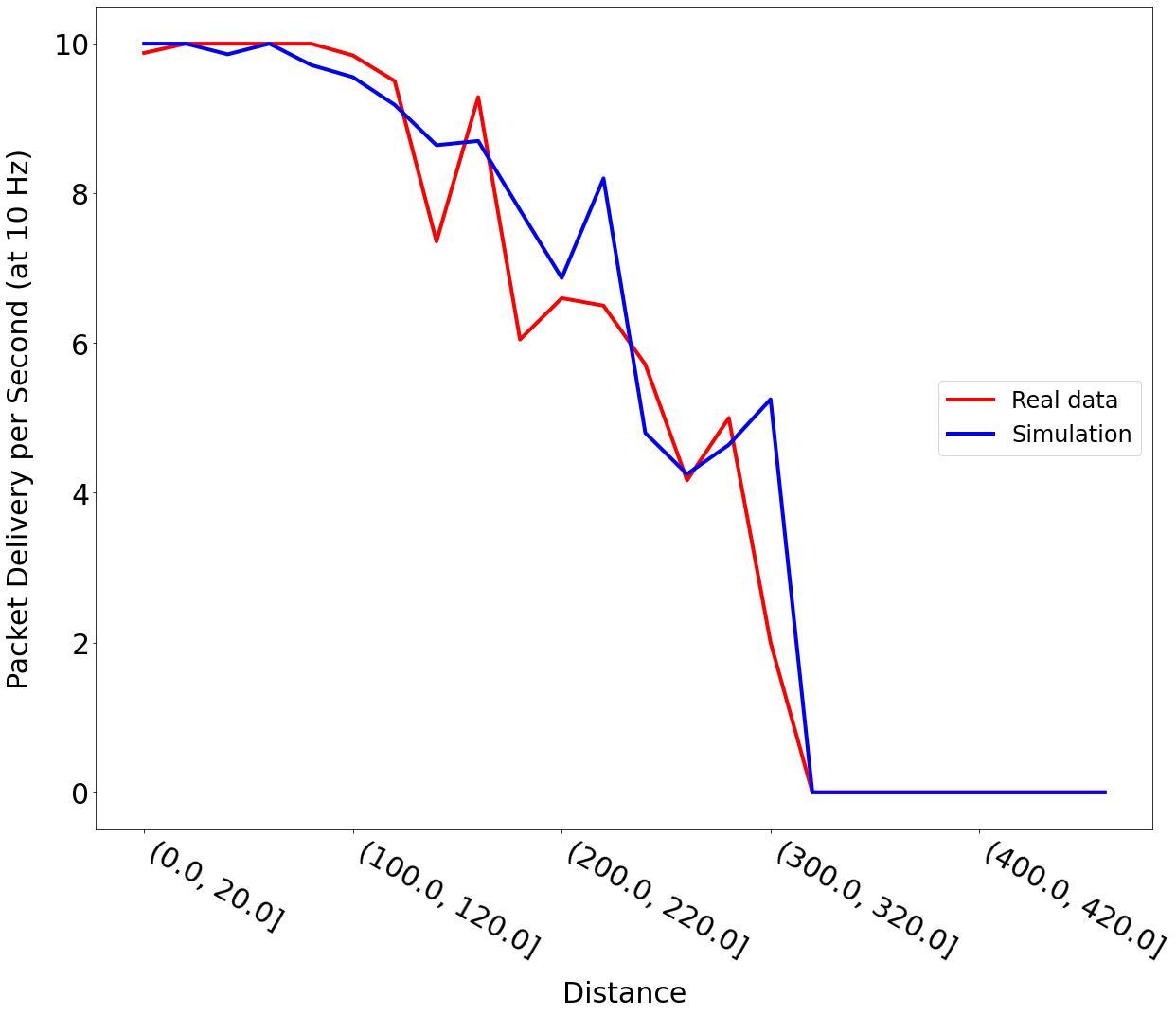}
             \end{tabular}
        \caption{Comparison of PDR vs distance for collected data vs simulation with default parameters (a), after introducing noise (b), and after running GA and calibrating the parameters (c)}
        \label{fig:per_curves}
\end{figure*}

With the built-in Veins example's default parameters, the model functions like a perfect channel that can send messages from anywhere without signal attenuation. The result was a significant RMSE of greater than 6.4, as shown in the left part of Figure \ref{fig:per_curves}. After increasing the noise from -110 dBm to -60 dBm the model performed better by having a delivery range close to the real data. This way we got an RMSE of 2.3 as shown in the middle part of Figure \ref{fig:per_curves}.

After that we defined the new pathloss model that uses Lognormal slow fading and Nakagami fast fading model and using the default parameters as shown in Table-\ref{tab:cal_params}, we got a PDR curve that is identical to the one in the middle of Figure \ref{fig:per_curves} (note that we fixed the noise here as its effects are known). That way, the default parameters just resemble an FSM channel. 

The Genetic Algorithm returned the set of calibrated parameters shown in Table-\ref{tab:cal_params} after more than 5,500 simulation runs. The calibrated model performed much better than all the previous models with an RMSE of about 0.908 and with PDR pattern that follow closely the testbed data PDR pattern as shown in the right part of Figure \ref{fig:per_curves} which demonstrates that the modified pathloss model provides a better approximation of the collected data after calibration.

\begin{figure*}[!tp]
        \begin{minipage}[t]{.45\linewidth}
            \centering
             \begin{tabular}{@{}c@{}}
  \includegraphics[width=3.1in]{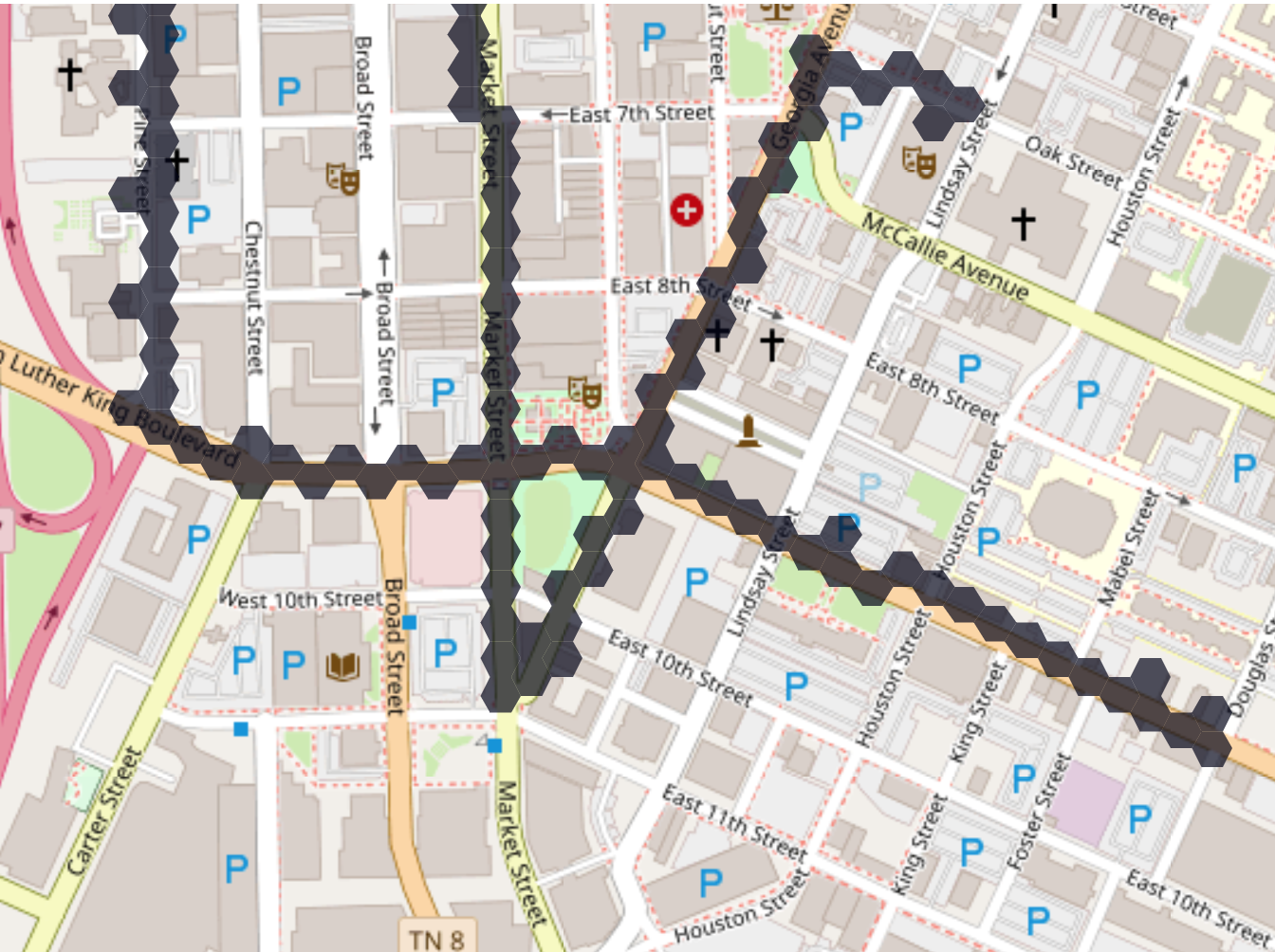}
  {$(a)$}
             \end{tabular}
             \begin{tabular}{@{}c@{}}
  \includegraphics[width=3.1in]{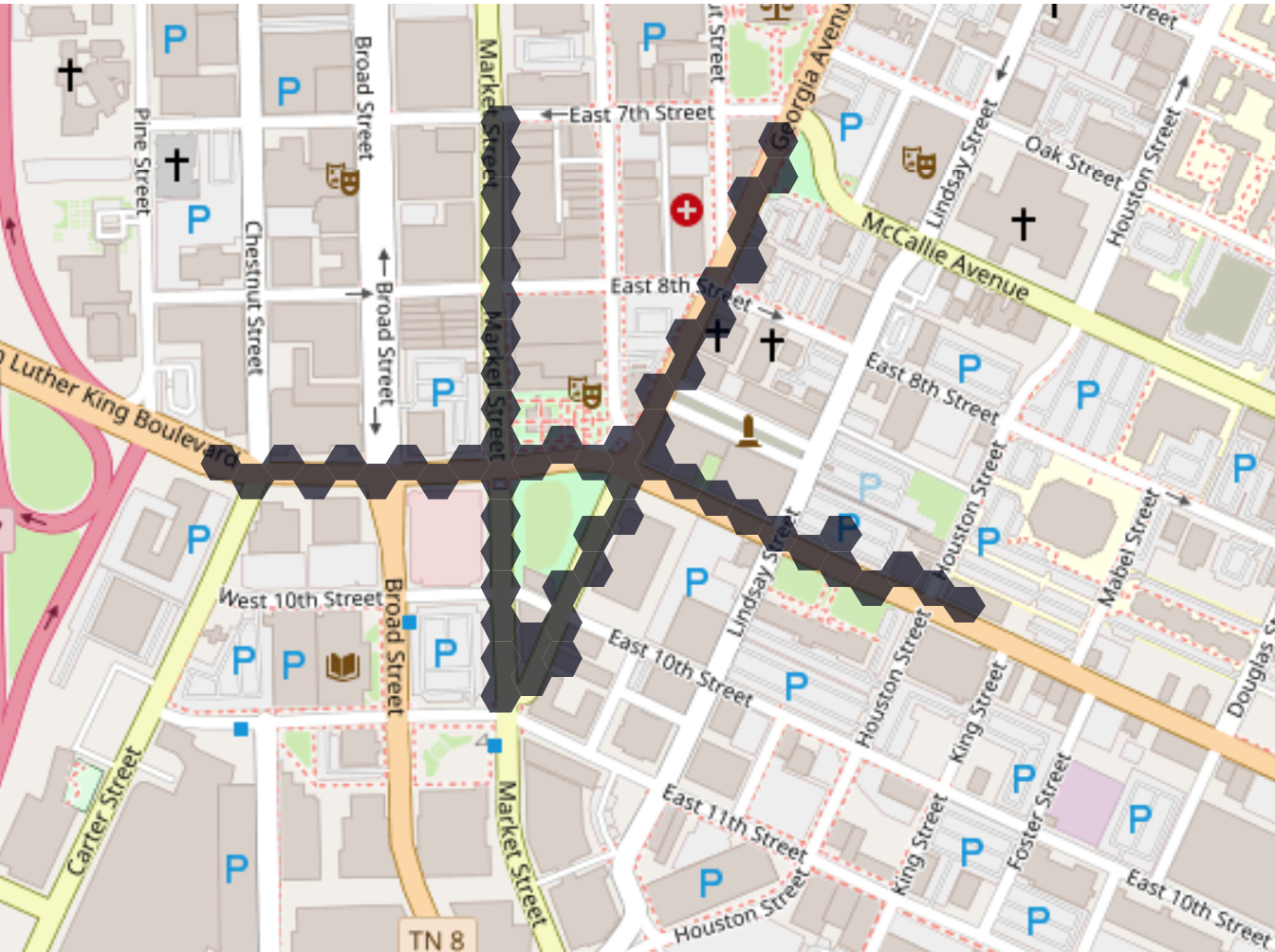}
  {$(b)$}
             \end{tabular}
        \end{minipage}
        \begin{minipage}[t]{.45\linewidth}
            \centering
             \begin{tabular}{@{}c@{}}
  \includegraphics[width=3.1in]{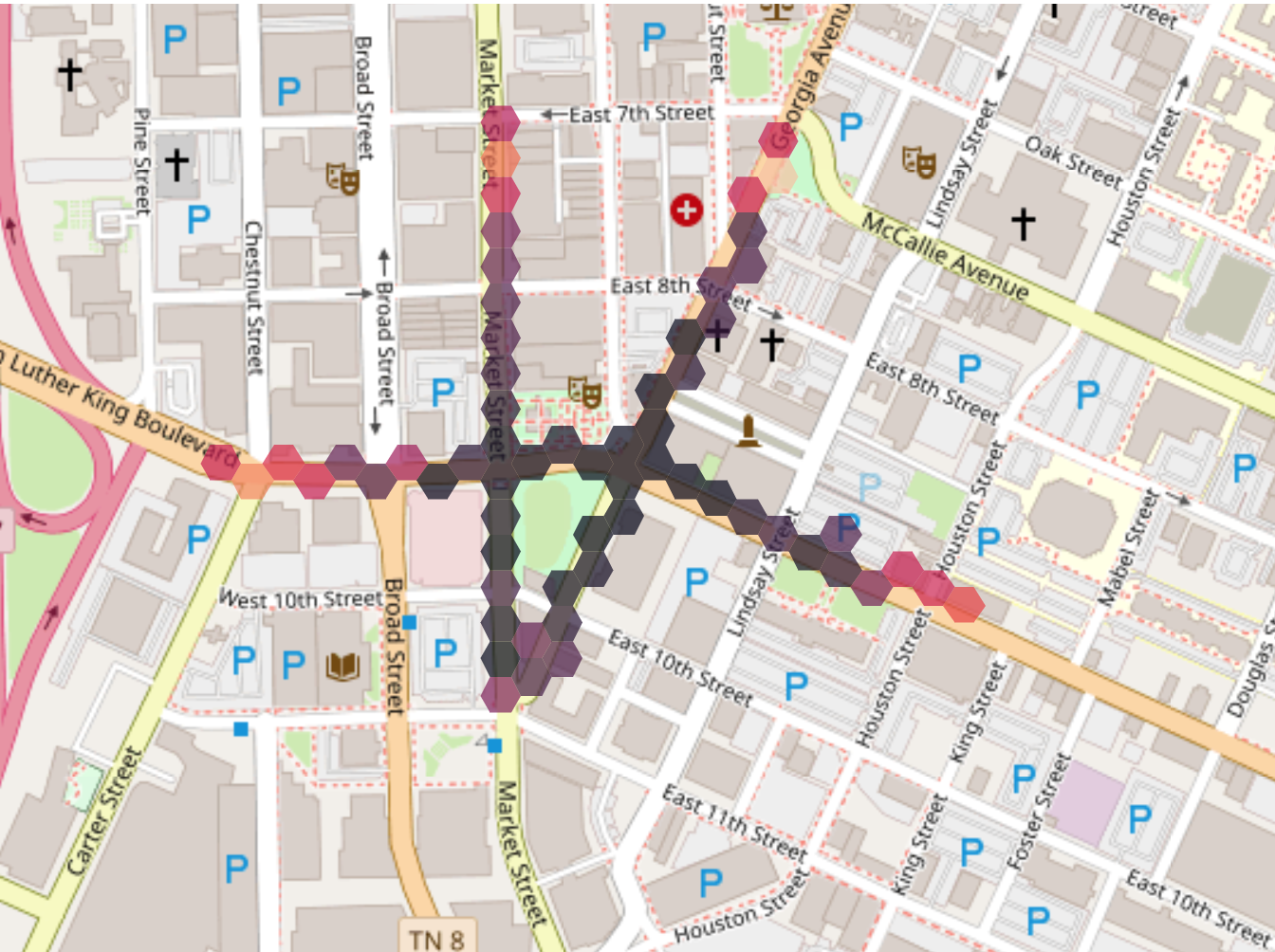}
  {$(c)$}
             \end{tabular}
             \begin{tabular}{@{}c@{}}
  \includegraphics[width=3.1in]{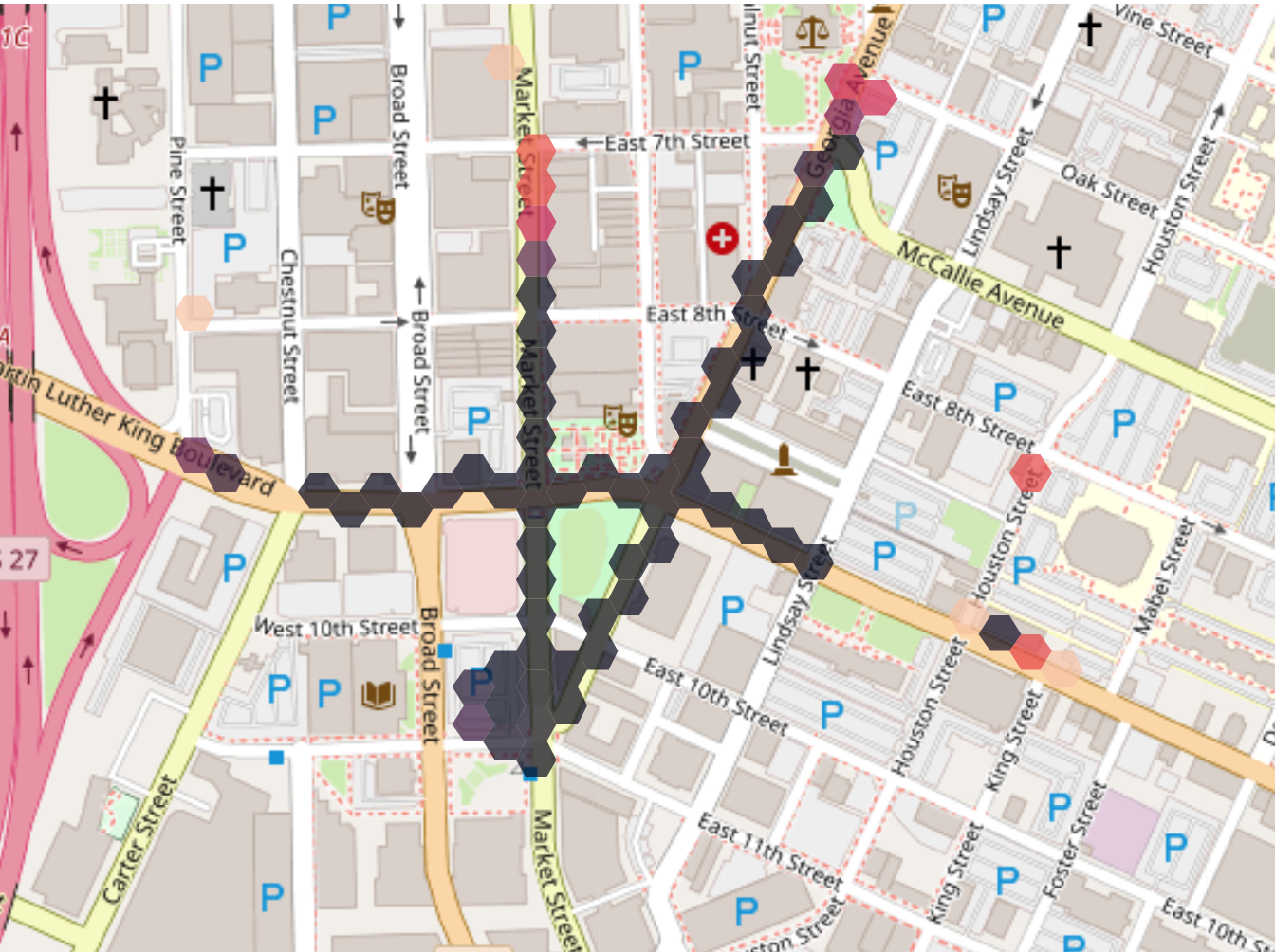}
  {$(d)$}
             \end{tabular}
        \end{minipage}
        \caption{Heatmaps of PDR under different simulation parameters. (a): with the built in values and with FSM pathloss. (b): after increasing the noise and introducing Lognormal and Nakagami models but before calibration. (c): after calibration using GA. (d): the real data from the testbed.}
        \label{fig:heatmaps}
\end{figure*}

Figure \ref{fig:heatmaps} shows the heatmaps that represents the packet delivery density at different locations of the real data and the different simulation runs (a, b, and c) compared to collected data (d). The heatmap of the real world data suggests some obstacles blocking the signal from the OBU. Modelling the obstacles isn't captured with the modified pathloss.


\begin{table}[!ht]
	\caption{The Target Parameters of this Study}\label{tab:cal_params}
	\begin{center}
		\begin{tabular}{l l l l l}
			ID & Parameter name & Range & Default value & Calibrated Value \\\hline
			1   & Transmission power & 20 - 40 mW & 20 mW & 30.16 mW \\
			2   & Data rate & 6, 12, 18, 27 Mbps & 6 Mbps & 18 Mbps\\
			3   & Noise & -110 dBm to -90 dBm  & -110 dBm & -90 dBm \\
			4   & Receiving Sensitivity & -120 dBm to -90 dBm  & -110 dBm & -114 dBm \\
			5   & Slow fading model & Free-space model (FSM) & FSM & Lognormal Model \\
			   &  & Lognormal Model      &  &  \\
			 &  & Nakagami Model &  & \\
			6   & Fast fading model & Lognormal Model (LNM)  & - & Nakagami \\
			   &  & Nakagami Model  &  \\
			7 & Alpha (Path Loss exponent of FSM) & 1.0 - 3.0 & 1.0 & 1.51 \\
			8 & System loss (of FSM) & 0.0 - 3.0 & 0.0 & 0.13\\
			9 & Sigma (Standard deviation of LNM) & 1.0 - 10.0 & 2.0 & 6.03 \\
			10 & Shape Factor (of Nakagami) & 1.0 - 3.5 & 1.0 & 2.0 \\
		\end{tabular}
	\end{center}
\end{table}

\section{VI Conclusion}

 In this paper we presented a novel process to calibrate VANET simulator channel model using real-world wireless communication data from the MLK Smart Corridor, an urban testbed in Chattanooga, TN. Veins was chosen as the target of our study. We incrementally built the model starting with an ideal channel that experiences no packet drops. We started by increasing the noise and then used a complex path loss.  We then implemented a complex pathloss that uses Lognormal slow and Nakagami fast fading models as currently Veins (and INET) allows only the freespace model for slow fading. We showed that the calibrated channel model performs better for DSRC channels compared compared to uncalibrated models in approximating the PDR curve of the field collected data.

 However, this study have not included the effects of shadowing by buildings and other vehicles. Also, multi-source noise and multi-path interference need to be further investigated. In the future work, more comparisons will include the raw data without calculating trends. That is, location, speed and heading will be included in the pathloss calculation. The study suggests more flexible approach in cascading slow- and fast- fading models in future releases of Veins and other VANET simulators. This will facilitate research in channel modelling. 

\section{VII Acknowledgement}

This work is partially supported by the National Science Foundation under Award CCRI-2120358 and U.S. Department of Energy’s Office of Energy Efficiency and Renewable Energy (EERE) under the Award Number DE-EE0009208. This report was prepared as an account of work sponsored by an agency of the United States Government. Neither the United States Government nor any agency thereof, nor any of their employees, makes any warranty, express or implied, or assumes any legal liability or responsibility for the accuracy, completeness, or usefulness of any information, apparatus, product, or process disclosed, or represents that its use would not infringe privately owned rights. Reference herein to any specific commercial product, process, or service by trade name, trademark, manufacturer, or otherwise does not necessarily constitute or imply its endorsement, recommendation, or favoring by the United States Government or any agency thereof. The views and opinions of authors expressed herein do not necessarily state or reflect those of the United States Government or any agency thereof.

\bibliographystyle{unsrt}  


\begin{thebibliography}{1}

\bibitem{sommer2019veins}
Christoph Sommer, David Eckhoff, Alexander Brummer, Dominik S Buse, Florian Hagenauer, Stefan Joerer, and Michele Segata.
\newblock Veins: The open source vehicular network simulation framework.
\newblock In {\em Recent advances in network simulation}, pages 215--252. Springer, 2019.

\bibitem{wegener2008traci}
Axel Wegener, Micha{\l} Pi{\'o}rkowski, Maxim Raya, Horst Hellbr{\"u}ck, Stefan Fischer, and Jean-Pierre Hubaux.
\newblock TraCI: an interface for coupling road traffic and network simulators.
\newblock In {\em Proceedings of the 11th communications and networking simulation symposium}, pages 155--163, 2008.

\bibitem{sommer2010bidirectionally}
Christoph Sommer, Reinhard German, and Falko Dressler.
\newblock Bidirectionally coupled network and road traffic simulation for improved IVC analysis.
\newblock {\em IEEE Transactions on mobile computing}, 10(1):3--15, 2010.

\bibitem{weber2021vanet}
Julia Silva Weber, Miguel Neves, and Tiago Ferreto.
\newblock VANET simulators: an updated review.
\newblock {\em Journal of the Brazilian Computer Society}, 27(1):1--31, 2021.

\bibitem{miloslavov2011validation}
Adelin Miloslavov, Malathi Veeraraghavan, and Brian L Smith.
\newblock Validation of vehicular network simulation models with field-test measurements.
\newblock In {\em 2011 7th International Wireless Communications and Mobile Computing Conference}, pages 853--859. IEEE, 2011.

\bibitem{su2016calibrating}
Peng Su, Joyoung Lee, and Byungkyu Brian Park.
\newblock Calibrating communication simulator for connected vehicle applications.
\newblock {\em Journal of Intelligent Transportation Systems}, 20(1):55--65, 2016.

\bibitem{mavromatis2017agile}
Ioannis Mavromatis, Andrea Tassi, Robert J Piechocki, and Andrew Nix.
\newblock Agile calibration process of full-stack simulation frameworks for V2X communications.
\newblock In {\em 2017 IEEE Vehicular Networking Conference (VNC)}, pages 89--96. IEEE, 2017.

\bibitem{adrian2018study}
Ronald Adrian, Selo Sulistyo, and I Wayan Mustika.
\newblock A study on communication system in VANET.
\newblock In {\em 2018 4th International Conference on Science and Technology (ICST)}, pages 1--6. IEEE, 2018.

\bibitem{rappaport1996wireless}
Theodore S Rappaport and others.
\newblock Wireless communications: principles and practice.
\newblock {\em prentice hall PTR New Jersey}, 2, 1996.

\bibitem{fall2012ns}
K Fall and V Kannan.
\newblock The ns Manual, May, 9, 2010.
\newblock 2012.

\bibitem{meszaros2019inet}
Levente M{\'e}sz{\'a}ros, Andras Varga, and Michael Kirsche.
\newblock Inet framework.
\newblock In {\em Recent Advances in Network Simulation}, pages 55--106. Springer, 2019.

\bibitem{torrent2004broadcast}
Marc Torrent-Moreno, Daniel Jiang, and Hannes Hartenstein.
\newblock Broadcast reception rates and effects of priority access in 802.11-based vehicular ad-hoc networks.
\newblock In {\em Proceedings of the 1st ACM international workshop on Vehicular ad hoc networks}, pages 10--18, 2004.

\bibitem{yin2006dsrc}
Jijun Yin, Gavin Holland, Tamer Elbatt, Fan Bai, and Hariharan Krishnan.
\newblock DSRC channel fading analysis from empirical measurement.
\newblock In {\em 2006 First International Conference on Communications and Networking in China}, pages 1--5. IEEE, 2006.

\bibitem{4291825}
George K. Karagiannidis, Nikos C. Sagias, and P. Takis Mathiopoulos.
\newblock $N{\ast}$Nakagami: A Novel Stochastic Model for Cascaded Fading Channels.
\newblock {\em IEEE Transactions on Communications}, 55(8):1453--1458, 2007.

\bibitem{9006382}
Austin Harris, Jose Stovall, and Mina Sartipi.
\newblock MLK Smart Corridor: An Urban Testbed for Smart City Applications.
\newblock In {\em 2019 IEEE International Conference on Big Data (Big Data)}, pages 3506--3511, 2019.

\bibitem{mitchell1995genetic}
Melanie Mitchell.
\newblock Genetic algorithms: An overview.
\newblock In {\em Complex.}, 1(1):31--39, 1995.

\bibitem{9449197}
Claude Nawej, Pius Owolawi, and Tom Walingo.
\newblock Design and Simulation of VANETs Testbed Using OpenStreetMap, SUMO, and NS-2.
\newblock In {\em 2021 IEEE 6th International Conference on Computer and Communication Systems (ICCCS)}, pages 582--587, 2021.

\bibitem{wang2009nctuns}
Shie-Yuan Wang and Chih-Liang Chou.
\newblock NCTUns simulator for wireless vehicular ad hoc network research.
\newblock {\em Ad Hoc Networks: New Research}, pages 97--123, 2009.

\bibitem{wang2007design}
Shie-Yuan Wang, CL Chou, and CC Lin.
\newblock The design and implementation of the NCTUns network simulation engine.
\newblock {\em Simulation Modelling Practice and Theory}, 15(1):57--81, 2007.

\bibitem{federal2002amendment}
Federal Communications Commission and others.
\newblock Amendment of the commissions rules regarding dedicated short-range communication services in the 5.850-5.925 ghz band (5.9 ghz band.
\newblock {\em FCC 02-302, https://docs. fcc. gov/public/attachments/FCC-02-302A1. doc}, 2002.

\bibitem{ieee2010802}
IEEE Standards Association, Z and others.
\newblock 802.11 p-2010-IEEE standard for information technology-local and metropolitan area networks-specific requirements-part 11: Wireless lan medium access control (mac) and physical layer (phy) specifications amendment 6: Wireless access in vehicular environments.
\newblock {\em URL http://standards. ieee. org/findstds/standard/802.11 p-2010. html}, 2010.

\bibitem{6998915}
IEEE Std 1609.3-2010/Cor 2-2014 (Corrigendum to IEEE Std 1609.3-2010).
\newblock IEEE Standard for Wireless Access in Vehicular Environments (WAVE) -- Network Systems Corrigendum 2: Miscellaneous Corrections.
\newblock {\em IEEE}, pages 1--33, 2014.

\bibitem{al2014comprehensive}
Saif Al-Sultan, Moath M Al-Doori, Ali H Al-Bayatti, and Hussien Zedan.
\newblock A comprehensive survey on vehicular ad hoc network.
\newblock {\em Journal of network and computer applications}, 37:380--392, 2014.

\bibitem{ganeshkumar2021obu}
N Ganeshkumar and Sanjay Kumar.
\newblock Obu (on-board unit) wireless devices in vanet (s) for effective communication—A review.
\newblock {\em Computational Methods and Data Engineering}, pages 191--202, 2021.

\bibitem{varga2010omnet++}
Andras Varga.
\newblock OMNeT++.
\newblock In {\em Modeling and tools for network simulation}, pages 35--59. Springer, 2010.

\bibitem{carneiro2010ns}
Gustavo Carneiro.
\newblock NS-3: Network simulator 3.
\newblock In {\em UTM Lab Meeting April}, 20:4--5, 2010.

\bibitem{issariyakul2009introduction}
Teerawat Issariyakul and Ekram Hossain.
\newblock Introduction to network simulator 2 (NS2).
\newblock In {\em Introduction to network simulator NS2}, pages 1--18. Springer, 2009.

\bibitem{SUMO2018}
Pablo Alvarez Lopez, Michael Behrisch, Laura Bieker-Walz, Jakob Erdmann, Yun-Pang Fl{\"o}tter{\"o}d, Robert Hilbrich, Leonhard L{\"u}cken, Johannes Rummel, Peter Wagner, and Evamarie Wie{\ss}ner.
\newblock Microscopic Traffic Simulation using SUMO.
\newblock In {\em The 21st IEEE International Conference on Intelligent Transportation Systems}. IEEE, 2018.

\bibitem{krajzewicz2012recent}
Daniel Krajzewicz, Jakob Erdmann, Michael Behrisch, and Laura Bieker.
\newblock Recent development and applications of SUMO-Simulation of Urban MObility.
\newblock {\em International journal on advances in systems and measurements}, 5(3\&4), 2012.

\bibitem{lopez2018microscopic}
Pablo Alvarez Lopez, Michael Behrisch, Laura Bieker-Walz, Jakob Erdmann, Yun-Pang Fl{\"o}tter{\"o}d, Robert Hilbrich, Leonhard L{\"u}cken, Johannes Rummel, Peter Wagner, and Evamarie Wie{\ss}ner.
\newblock Microscopic traffic simulation using sumo.
\newblock In {\em 2018 21st international conference on intelligent transportation systems (ITSC)}, pages 2575--2582. IEEE, 2018.

\bibitem{lim2017sumo}
Kit Guan Lim, Chun Hoe Lee, Renee Ka Yin Chin, Kiam Beng Yeo, and Kenneth Tze Kin Teo.
\newblock SUMO enhancement for vehicular ad hoc network (VANET) simulation.
\newblock In {\em 2017 IEEE 2nd international conference on automatic control and intelligent systems (I2CACIS)}, pages 86--91. IEEE, 2017.

\bibitem{fellendorf2010microscopic}
Martin Fellendorf and Peter Vortisch.
\newblock Microscopic traffic flow simulator VISSIM.
\newblock In {\em Fundamentals of traffic simulation}, pages 63--93. Springer, 2010.

\bibitem{homepage2022omnet++}
OMNET++ homepage.
\newblock {\em \url{https://omnetpp.org/intro/}}, Accessed: 2022-08-01.

\end{thebibliography}

\end{document}